\documentclass[fleqn,twoside]{article}%
\usepackage{amsmath}
\usepackage{amsfonts}
\usepackage{amssymb}
\usepackage{graphicx}
\usepackage{cite}
\def\be{\begin{equation}}
\def\ee{\end{equation}}
\def\bi{\bibitem}
\begin{document}
\title{Bianchi type I cosmological model with a viscous fluid.}
\author{A. Banerjee$^1$, S. B. Duttachoudhury and Abhik Kumar Sanyal$^2$}
\maketitle
\noindent
\begin{center}
\noindent
Dept of Physics, Jadavpur University, Calcutta 700 032, India.\\
\end{center}
\footnotetext{\noindent
Electronic address:\\
\noindent
$^1$ asit@juphys.ernet.in\\
$^2$ sanyal\_ ak@yahoo.com;
Present address: Dept. of Physics, Jangipur College, India - 742213.}
\noindent
\abstract{Bianchi I cosmological models consisting of a fluid with both bulk and shear viscosity are studied. It is shown how the dynamical importance of the shear and the fluid density change in the course of evolution. Exact solutions with an equation of state $\rho = p$ for a stiff fluid are also obtained in several special cases, assuming the viscosity coefficients to be the power functions of the density. The results are, in some relevant cases, compared with those of Belinski$\check{\mathrm{i}}$ and Khalatnikov (1976) in the asymptotic limits and are seen to agree with them in that the models start with $\rho = 0$ at the beginning and evolve with the creation of matter by the gravitational field, finally approaching the Friedmann universe.}
\maketitle
\flushbottom
\section{Introduction:}
The presence of viscosity in the fluid content introduces many interesting features in the dynamics of homogeneous cosmological models \cite{1, 2, 3, 4, 5}. The dissipative mechanisms not only modify the nature of the singularity usually occurring for a perfect fluid, but also can successfully account for the large entropy per baryon in the present universe. Misner \cite{6, 7} suggested that any anisotropy in an expanding universe would be reduced to a rather insignificant level today by neutrino viscosity. Murphy \cite{8} presented an exactly soluble cosmological model of Friedmann type in the presence of bulk viscosity alone. Later, Banerjee and Santos \cite{9} extended the calculations to more general cases such as $k \pm 1$. It was shown, however, that in all the cases where the singularity is said to appear at infinite past, the fluid did not satisfy Hawking-Penrose energy conditions.\\

Exact solutions for homogeneous anisotropic models are not much known in the literature. There are, however, a few \cite{10, 11} which utilize certain simplifying assumptions to get exact solutions at the cost of a physically reasonable equation of state. Belinski$\check{\mathrm{i}}$ and Khalatnikov \cite{4} assumed an equation of state of the form $\rho \propto p$, but did not give any exact solution. They have, however, investigated some general features of the isotropic and anisotropic homogeneous cosmological models in the presence of bulk as well as shear viscosity in asymptotic limits. We consider in this paper the Bianchi I model with a fluid characterized by both bulk and shear viscosity and having an equation of state $\rho \propto p$. The viscosity coefficients are further assumed to be power functions of the matter density as suggested by Belinski$\check{\mathrm{i}}$ and Khalatnikov \cite{4}. Exact solutions in several particular cases for stiff fluid $\rho = p$ are worked out.\\

In Sec. II we have considered Einstein's field equation for a Bianchi I cosmological model and discussed how the dynamical importance of the shear and matter change in the course of the cosmological evolution. The entropy variation is also explicitly stated. In Sec. III exact solutions are obtained and their asymptotic characteristics are studied. The general behavior in the limits is compared with that of Belinski$\check{\mathrm{i}}$ and Khalatnikov \cite{4} at some places. Explicit solutions could be obtained considering the viscous coefficients only for a few restricted power functions of the mass density. These include the special cases with constant viscosity coefficients. In many of these cases the matter density vanishes at the initial instant, then increases in the course of evolution and finally reduces to zero. In such cases, therefore, at the initial singularity the metric is determined by the free space Einstein equations. In this context Belinski$\check{\mathrm{i}}$ and Khalatnikov \cite{4} remarked that the gravitational field creates the matter in the course of evolution.

\section{Einstein's field equations and some general results}
The metric of the homogeneous Bianchi type I model is
\be\label{2.1} ds^2 = -dt^2 + e^{2\alpha} dx^2 + + e^{2\beta} dy^2 + e^{2\gamma} dz^2,\ee
where $\alpha,~\beta~ \mathrm{and}~ \gamma$ are functions of time alone. The energy momentum tensor of the viscous fluid \cite{12} is given by
\be\label{2.2} T_{ij} = (\rho+\bar{p})v_iv_j + \bar{p} g_{ij} - \eta \mu_{ij},\ee
with
\be\label{2.3} \bar{p} = p - \left(\zeta - {2\over 3}\eta\right)\theta,\ee
and
\be\label{2.4} \mu_{ij} = v_{i;j} + v_{j;i} + v_i v^{a}v_{j;a} + v_j v^{a}v_{i;a},\ee
where,
\be \theta = {v^a}_{;a}.\ee
In the above equations $\zeta ~\mathrm{and}~ \eta$ stand for the bulk and shear viscosity coefficients, $\rho$ and $p$ are the mass density and pressure, respectively, $\bar{p}$ is the effective pressure, and $v_i$ represents the four-velocity, so that
\be \label{2.5} v_i v^j = -1.\ee
Choosing units $8\pi G = C = 1$, Einstein's field equations can be written as
\be \label{2.6} {R^i}_j - {1\over 2}{\delta^i}_j R = -{T^i}_j.\ee
Using comoving coordinates, so that $v_i = {\delta_0}^i$, the explicit
forms of Eq. \eqref{2.6} are
\be \label{2.7} {9\over 2}\left({\dot R\over R}\right)^2 - {1\over 2}(\dot\alpha^2 + \dot \beta^2 + \dot\gamma^2) = \rho,\ee
\be \label{2.8}\ddot \beta + \ddot\gamma + {3\over 2}\left({\dot R\over R}\right)(\dot\beta + \dot \gamma - \dot\alpha) + {1\over 2}(\dot\alpha^2 + \dot \beta^2 + \dot\gamma^2) = -(\bar p - 2\eta\dot\alpha),\ee
\be \label{2.9}\ddot\gamma + \ddot\alpha + {3\over 2}\left({\dot R\over R}\right)(\dot \gamma + \dot\alpha - \dot\beta)
+ {1\over 2}(\dot\alpha^2 + \dot \beta^2 + \dot\gamma^2)=-(\bar p - 2\eta\dot\beta),\ee
\be \label{2.10}\ddot\alpha + \ddot\beta + {3\over 2}\left({\dot R\over R}\right)(\dot\alpha + \dot\beta - \dot \gamma )
+ {1\over 2}(\dot\alpha^2 + \dot \beta^2 + \dot\gamma^2)=-(\bar p - 2\eta\dot\gamma),\ee
where the dot indicates time differentiation and
\be\label{2.11} R^3 = \exp{(\alpha + \beta + \gamma}).\ee
The usual definitions of the dynamical scalars such as the expansion scalar $\theta$ and the shear scalar $\sigma$ are considered to be
\be\label{2.12} \theta = {v^i}_{;i} ~~~~~\mathrm{and}~~~~~\sigma^2 = {1\over 2}\sigma_{ij}\sigma^{ij},\ee
where
\be\label{2.13} \sigma_{ij}= v_{i;j} + {1\over 2}(v_{i;k} v^k v_j + v_{j;k} v^k v_i) + {1\over 3} \theta(g_{ij} + v_i v_j).\ee
For the Bianchi type I metric with comoving coordinates we have
\be\label{2.14} \theta = 3\left(\dot R\over R\right),\ee
and
\be\label{2.15} 2\sigma^2 = (\dot\alpha^2 + \dot\beta^2 + \dot\gamma^2) - {1\over 3}\theta^2.\ee
The field equations \eqref{2.7} - \eqref{2.10} now yield
\be\label{2.16} {T^4}_4 = {1\over 3} \theta^2 - \sigma^2 = \rho,\ee
and
\be\label{2.17}g_{ij} G^{ij} = 2\dot\theta + {4\over 3}\theta^2
+ 2\sigma^2 = \rho - 3(\bar p - \zeta\theta).\ee
One can further obtain from the Bianchi identity
\be\label{2.18} \dot \rho = -(\rho + p)\theta + \zeta \theta^2 + 4\eta\sigma^2.\ee
It follows directly from Eq. \eqref{2.18} that for contraction, that is, $\theta < 0$ we have $\dot\rho > 0$ so that the matter density increases or decreases depending on whether the viscous heating is greater or less than the cooling due to expansion. It may be mentioned here that for an ultrarelativistic fluid $\rho = {1\over 3}p$ and $\zeta = 0$, Stewart \cite{13} has shown that the rate of viscous heating does not exceed one-half the rate of adiabatic cooling due to expansion.\\

Now, eliminating $\sigma^2$ from Eqs. \eqref{2.17} and \eqref{2.18}, one readily obtains
\be\label{2.19} \theta = {3\over 2}(\rho - p + \zeta\theta) - \theta^2,\ee
and
\be\label{2.20} \dot \rho = - (\rho + p)\theta + \zeta\theta^2 + 4\eta \left({1\over 3}\theta^2 - \rho\right).\ee
The relation \eqref{2.19} may be written as
\be\label{2.21} {d\over dt}\left[\ln{(\theta^2 R^6)}\right] = 3\left[(\rho - p)\theta^{-1} + \zeta\right].\ee
Using Eq. \eqref{2.16}, Eq. \eqref{2.19} can also be expressed in a different
form such as
\be\label{2.22} \dot\theta = -2\sigma^2 -{1\over 3}\theta^2 - {1\over 2}\left[\rho + 3(\rho - \zeta \theta)\right].\ee
This is exactly the Raychaudhuri equation \cite{14}. Further, we have
\be\label{2.23} R_{ij} v^i v^j = -{1\over 2}[\rho + 3(p - \zeta \theta)].\ee
The Hawking-Penrose energy condition is satisfied when $R_{ij}v^i v^j \le 0$. Thus, in a contracting model, so long as the fluid density and pressure remain positive, the energy conditions are satisfied. When the bulk viscosity is insignificant the energy condition is independent of the viscosity of the fluid. Again from the Raychaudhuri equation \eqref{2.22} it is evident that with the energy condition being satisfied, $\dot\theta < 0$, so that there is no bounce from a minimum volume.\\

Now for $\theta \ne 0$, that is, for a nonstatic model, one has, in view of Eq. \eqref{2.16},
\be\label{2.24} \left({\sigma^2\over \theta^2}\right)^{\textbf{.}} = - \left({\rho\over \theta^2}\right)^{\textbf{.}}\ee
Using the expressions for $\dot\rho ~\mathrm{and}~ \dot \theta$ from Eqs. \eqref{2.19} and \eqref{2.20} in Eq. \eqref{2.24}, one obtains after simplification the following result:
\be\label{2.25} \left({\sigma^2\over \theta^2}\right)^{\textbf{.}} = - \left({\rho\over \theta^2}\right)^{\textbf{.}} = - \left({\sigma^2\over \theta^2}\right)\left[3(\rho - p)\theta^{-1} + 3\zeta + 4\eta\right].\ee
The reasonable physical properties of the fluid demand $\rho \ge p > 0,~ \zeta > 0,~ \mathrm{and} ~ \eta > 0$, so that for an expanding model $(\theta > 0)$, we have $\left({\rho\over \theta^2}\right)^{\textbf{.}} > 0$ and $\left({\sigma^2\over\theta^2}\right)^{\textbf{.}} < 0$. It is evident, therefore, that $\left({\rho\over \theta^2}\right)$ increases with time, while $\left({\sigma^2\over\theta^2}\right)$ decreases. The dynamical influence of matter, therefore, increases with expansion, whereas that of shear decreases. For contraction, however, $\theta < 0$ and nothing can be said with certainty. It is interesting to note that for a stiff fluid, that is, $\rho = p$, $\left({\rho\over\theta^2}\right)$ is greater than zero and $\left({\sigma^2\over\theta^2}\right)$ is less than zero, so long as viscosity coefficients are positive, and this behavior holds irrespective of whether the model expands or contracts. Combining Eqs. \eqref{2.21} and \eqref{2.25}, one obtains
\be\label{2.26} \left[\ln{(\sigma^2 R^6)}\right] = - 4\eta,\ee
which in turn can also be written as
\be\label{2.27} (\sigma^2)^{\textbf{.}} = -2(2\eta + \theta)\sigma^2.\ee
This is the shear propagation equation. It follows from Eq. \eqref{2.27} that for the expanding model $\theta > 0$ the shear decreases with time. The rate of work done by anisotropic stresses augments the shear dissipation. It is also evident from Eq. \eqref{2.27} that the shear dissipation depends on the expansion rate $(\theta)$, which is again affected by the presence of bulk viscosity as is evident from Eq. \eqref{2.22}. The bulk viscosity has thus a significant role in the process of the shear dissipation mechanism.\\

When $\eta$ is assumed to be a constant the relation \eqref{2.27} can be directly integrated to yield
\be\label{2.28} \sigma^2 = \left({\sigma_0^2 \over R^6}\right)e^{-4\eta t},\ee
$\sigma_0$ being the integration constant. The effect of shear viscosity is to reduce the anisotropy in the course of time in the form of an exponential factor. This purely relativistic result is due to Misner \cite{6, 7}.\\

Following Belinski$\check{\mathrm{i}}$ and Khalatnikov \cite{4} the time derivative of the entropy density in the model is given by
\be\label{2.29} {\dot\Sigma\over \Sigma} = {\dot\rho\over (\rho + p)},\ee
where $\Sigma$ is the entropy density. Defining the total entropy by $S = R^3\Sigma$, one gets from Eq. \eqref{2.20} using Eq. \eqref{2.29} the relation
\be\label{2.30} {\dot S\over S} = {(\zeta \theta^2 + 4\eta \sigma^2)\over (\rho+p)}.\ee
We now restrict ourselves to an equation of state of the form
\be\label{2.31} p = (\gamma - 1)\rho,~~~~1 \le \gamma \le 2,\ee
and assume that the viscosity coefficients are constants so that $\zeta = \zeta_0$ and $\eta = \eta_0$ The qualitative aspects of the presence of viscosity will, however, be present in such a restricted case also (see Misner \cite{6, 7} and Treciokas and Ellis \cite{15}. In this case Eq. \eqref{2.30} can be written as
\be\label{2.32} {\dot S\over S} = {\left[\zeta_0 + 4 \eta_0 \left({\sigma^2 \over \theta^2}\right)\right]\over \gamma\left({\rho\over \theta^2}\right)}.\ee
By further differentiation with respect to time and using Eqs. \eqref{2.25} and \eqref{2.32}, we find
\be\label{2.33} \begin{split}S^{-1} \ddot S =& \gamma^{-2} \left({\rho\over \theta^2}\right)^{-2}\bigg[\zeta_0^2 - \left({\sigma^2\over \theta^2}\right)\left({\rho\over \theta^2}\right) \times \left\{16\eta_0^2 + \gamma(2-\gamma)(3\zeta_0 + 4\eta_0)\theta \right\}\\&
+ 8\eta_0(\gamma-1)\left(\zeta_0 + {2\over 3}\eta_0\right)+ 3\gamma\zeta_0^2\bigg].\end{split}\ee
For an expanding model, $\left({\sigma^2\over \theta^2}\right)$ decreases with time, whereas $\left({\rho\over \theta^2}\right)$ increases, as was already discussed. The minimum of $\left({\sigma^2\over \theta^2}\right)$ is zero, when $\left({\rho\over \theta^2}\right) = {1\over 3}$ and
$S^{-1} \ddot S = \zeta_0^2\gamma^{-2} \left({\rho\over \theta^2}\right)^{-2}$, which is greater than zero. In the
course of time $\left({\rho\over \theta^2}\right)$ decreases and we consider the extreme
case when $\left({\rho\over \theta^2}\right) = 0$. At this instant $\left({\sigma^2\over \theta^2}\right) = {1\over 3}$ and from Eq. \eqref{2.33} it is evident that $\ddot S < 0$. The relation \eqref{2.32} indicates that $\dot S > 0$, that is, the total entropy always increases for nonnegative values of matter density, whereas for expansion, $\ddot S$ is initially negative and later becomes positive in the course of time. Since ${\dot S\over S} \rightarrow \infty$ as $\left({\rho\over \theta^2}\right) \rightarrow 0$, we have the $S - t$ curve intersecting the time axis. It means that $S$ reduces to zero at some finite time. The picture is more clear for a stiff fluid when $\rho = p$. We have then from Eq. \eqref{2.25}
\be {\left({\sigma^2\over \theta^2}\right)^{\textbf{.}}\over \left({\sigma^2\over \theta^2}\right)} = (3\zeta_0 + 4\eta_0),\ee
which yields on integration
\be \label{2.34} \left({\sigma^2\over \theta^2}\right) = A^2 e^{-{(3\zeta_0 + 4\eta_0)t}},\ee
where $A^2$ is the magnitude of the ratio ${\sigma^2\over \theta^2}$ at $t = 0$. It is evident from Eq. \eqref{2.34} that for both $\zeta$ and $\eta$ as constants the ratio of shear to expansion decays exponentially and the rate falls in the absence of either bulk or shear viscosity. From Eqs. \eqref{2.16} and \eqref{2.34} one gets the expression for ${\rho\over \theta^2}$ in the form
\be\label{2.35} {\rho\over \theta^2} = {1\over 3} - A^2 e^{-(3\zeta_0 + 4\eta_0)t}.\ee
So, initially, if one starts with zero mass density at some
finite time one must have ${\left(\rho\over \theta^2\right)_{max}} = {1\over 3}$ at $t \rightarrow 0$. In view of Eqs. \eqref{2.34} and \eqref{2.35}, Eq. \eqref{2.32} can now be written as
\be\label{2.36} {\dot S\over S} = {\zeta_0 + 4\eta_0 A^2 e^{-(3\zeta_0 + 4\eta_0)t}\over 2\left({1\over 3} - A^2 e^{-(3\zeta_0 + 4\eta_0)t}\right)},\ee
which in turn yields on integration
\be\label{2.37} S = S_0\left[{1\over 3}e^{3\zeta_0 t} - A^2 e^{-4\eta_0 t}\right]^{1\over 2},\ee
with $S_0$ being the constant of integration. It is evident that at
\be\label{2.38} t = \left[\ln{(3A^2)}\right](3\zeta_0 + 4\eta_0)^{-1},\ee
the total entropy $S = 0$, when we also have ${\rho\over \theta^2} = 0$. Again as $t\rightarrow \infty,~ S\rightarrow \infty$ and ${\rho\over \theta^2}$ approaches its maximum value. Combining Eqs. \eqref{2.35} and \eqref{2.37} together one can also write
\be\label{2.39} S^2 = S_0^2 \left({\rho\over \theta^2}\right)e^{3\zeta_0 t},\ee
which in turn demands that the matter density must be nonnegative in this case.

\section{Special solutions for a Bianchi I model with a viscous fluid}

In what follows we consider some special cases with restrictions on the behavior of the bulk and shear viscosity coefficients. It is true that the assumptions regarding these viscosity coefficients may not always be valid in an actual fluid throughout the entire evolution history of the cosmological models hitherto discussed: the solutions are nevertheless interesting in indicating the role of viscosity in cosmological evolution.\\

In view of Eqs. \eqref{2.18} and \eqref{2.26} we now have the relation
\be\label{3.1} {d\over dt}\left[\rho + \sigma^2) R^6\right] = \zeta \theta^2 R^6.\ee
Using Eq. \eqref{2.16} in Eq. \eqref{3.1} we further obtain
\be\label{3.2} {d\over dt}\left[\ln{(\theta^2 R^6)}\right] = 3\zeta.\ee

\noindent
\textbf{Case - 1:}\\
In this case we consider $\zeta = 0$ and $\eta = \eta_0 \rho^n$, where $\eta_0$ and $n$ are constants. Equation \eqref{3.2} can immediately be integrated in this case to yield
\be\label{3.3} R^3 = {R_0}^3 t,\ee
where $R_0$ is an arbitrary constant and the time coordinate is
chosen such that the proper volume vanishes at $t = 0$. It is,
however, not difficult to show that one will get the same
solution \eqref{3.3} if one assumes the shear and bulk viscosity to
be absent for a perfect fluid. The expansion scalar $\theta$ is given
by
\be\label{3.4} \theta = t^{-1}.\ee
Equation \eqref{2.26} now yields in view of Eq. \eqref{2.16}
\be\label{3.5} \left({\sigma^2\over \theta^2}\right)^{-1}\left({\sigma^2\over \theta^2}\right)^\textbf{.}  = - 4\eta_0 \theta^{2n}\left({1\over 3} - {\sigma^2\over \theta^2}\right)^n.\ee
Writing ${\sigma^2\over \theta^2} = y$ and using the relation \eqref{3.4}, Eq. \eqref{3.5} can be written as a first-order differential equation
\be\label{3.6}{\dot y\over y}= -4\eta_0 t^{-2n}\left({1\over 3} - y\right)^n.\ee
In view of Eq. \eqref{2.16} $\rho$ is positive when $\left({1\over 3} - y\right) > 0$. For $n = 1$ the solution for $y$ is obtained by integrating Eq. \eqref{3.6} as
\be \label{3.7} y = {\sigma^2\over \theta^2} = {1\over 3}\left(1 + a^2 e^{-{4\eta_0\over 3t}}\right)^{-1},\ee
where $a^2$ is a positive constant. It follows from Eq. \eqref{3.7}, in view of Eq. \eqref{3.4}, that
\be \label{3.8} \sigma^2  = {1\over 3 t^2}\left(1 + a^2 e^{-{4\eta_0\over 3t}}\right)^{-1}.\ee
Using Eq. \eqref{2.16} we therefore obtain,
\be \label{3.9} \rho  = \left({a^2\over 3 t^2}\right)e^{-{4\eta_0\over 3t}}\left(1 + a^2 e^{-{4\eta_0\over 3t}}\right)^{-1}.\ee
Now consider an expanding model, for which $\theta > 0$. In this case at $t \rightarrow 0,~ R^3 \rightarrow 0$ and both the shear and expansion scalars attain infinitely large magnitudes, while the density reduces to zero. It is an interesting behaviour as noted previously by Belinski$\check{\mathrm{i}}$ and Khalatnikov \cite{4}. The singularity at $t = 0$ in this case corresponds to vanishing proper volume $R^3 = 0$. But, unlike in the usual case of cosmological singularity, the density vanishes in the limit instead of increasing to infinity. The matter density subsequently increases and again decreases to approach zero magnitude at the final stage at $t\rightarrow \infty$, as is evident from Eq. \eqref{3.9}. In this limit, however, both the expansion $(\theta)$ and shear $(\sigma^2)$ scalars vanish and $R^3 \rightarrow \infty$. In other words the model may be said to approach asymptotically the isotropic Friedmann universe (cf. Belinski$\check{\mathrm{i}}$ and Khalatnikov \cite{4}). On the other hand, collapse may be discussed for a negative value of $t$ when $t < 0,~\theta < 0$, which represents a contracting model. As $t\rightarrow -\infty$, all the quantities such as $\theta,~\sigma^2 ~\mathrm{and}~ \rho$ vanish with infinitely large proper volume representing a Friedmann model. In the course of time as $t \rightarrow 0$, the proper volume reduces to zero whereas $\theta,~\sigma^2 ~\mathrm{and}~ \rho$ all increase indefinitely. These properties in asymptotic limits only are discussed by Belinski$\check{\mathrm{i}}$ and Khalatnikov \cite{4}.\\

For $n = {3\over 2}$ the relation \eqref{3.2} takes the following form:
\be\label{3.10} {\dot y\over y} = -\left(4\eta_0\over t^3\right)\left({1\over 3} - y\right)^{3\over 2}.\ee
Since ${\left(\sigma^2\over \theta^2\right)^\textbf{.}} < 0$ for positive $\zeta$ and $\eta$, that is, $\dot y < 0$, $t$ should assume only positive values and thus one can have only expansion allowed in this case $(\theta > 0)$. Integrating Eq. \eqref{3.10}, we have
\be\label{3.11} \ln{\left[1-(1-3y)^{1\over 2}\over 1 + (1-3y)^{1\over 2}\right]} + {2\over (1-3y)^{1\over 2}} = {2\eta_0\over 3\sqrt 3}t^{-2} + \mathrm{constant}.\ee

For $n = 2$, we find from Eq. \eqref{3.6}, the solution of $y$ is given by
\be\label{3.12} 3\ln{\left[y\over \left({1\over 3} - y\right)\right]} + {1\over \left({1\over 3} - y\right)} = {4\over 9}\eta_0 t^{-3} + \mathrm{constant}.\ee
Though the solutions \eqref{3.11} and \eqref{3.12} are not in closed form, it is not very difficult to investigate the properties of these models at limits. The analysis is done in an identical manner as for $n = 1$. The behaviour can be seen to be almost identical in the limits $t\rightarrow 0$ or $t \rightarrow \infty$.\\

For $n = {1\over 2}$, Eq. \eqref{3.6} can be expressed as
\be\label{3.13} {\dot y\over y} = -\left(4\eta_0\over t\right)\left({1\over 3} - y\right)^{1\over 2},\ee
which on integration yields
\be \ln{\left[1-(1-3y)^{1\over 2}\over 1 + (1-3y)^{1\over 2}\right]}  = \left({t\over t_0}\right)^{-{4\eta_0\over \sqrt 3}}.\ee
where $t_0$ is the constant of integration and is less than $t$ as is clear from the above equation, which on further simplification gives
\be\label{3.14} y = {4\over 3}\left[{\left({t\over t_0}\right)^{-{4\eta_0\over \sqrt 3}}\over \left(1 + \left({t\over t_0}\right)^{-{4\eta_0\over \sqrt 3}}\right)^2}\right].\ee
One can now use Eq. \eqref{3.4} in Eq. \eqref{3.14} to obtain an expression for $\sigma^2$ (remembering, $y = \sigma^2/\theta^2$) as
\be\label{3.15} \sigma^2 = \left({4\over 3 t^2}\right)
\left({t\over t_0}\right)^{-{4\eta_0\over \sqrt 3}} \left(1 + \left({t\over t_0}\right)^{-{4\eta_0\over \sqrt 3}}\right)^{-2},\ee
and also
\be\label{3.16} \rho = \left({1\over 3t^2}\right)\left[1 - \left({t\over t_0}\right)^{-{4\eta_0\over \sqrt 3}}\right]^2 \left[1 + \left({t\over t_0}\right)^{-{4\eta_0\over \sqrt 3}}\right]^{-2}.\ee
For an expanding model $\theta > 0$. In this case, as $t\rightarrow t_0$, all the parameters $\theta,~R^3~\mathrm{and}~\sigma^2$ remain finite, while the matter density vanishes. On the other hand, as $t\rightarrow \infty$ the expansion $(\theta)$ and the shear $(\sigma^2)$ scalars vanish although the proper volume increases indefinitely, while the matter density $\rho$ approaches zero. Thus asymptotically the picture represents an isotropic Friedmann universe.\\

The simplest case is for $n = 0$, that is, $\eta = \eta_0$. We now have from Eq. \eqref{3.6}
\be\label{3.17} {\dot y\over y} = -4\eta_0,\ee
so that, $y = {\sigma^2\over \theta^2} \propto e^{-4\eta_0 t}$, yielding the relation
\be \label{3.18} \sigma^2 = \left({\sigma_0^2\over t^2}\right)e^{-4\eta_0 t}.\ee Further,
\be \label{3.19} \rho = {1\over 3}\theta^2 - \sigma^2 = {1\over t^2}\left[{1\over 3} - \sigma_0^2 e^{-4\eta_0 t} \right].\ee
The behavior in this case is quite different from the previous cases at least in the initial phase of expansion. Here at $t\rightarrow 0$, the physical and geometrical quantities such as $\rho,~\theta,~\sigma^2$ become infinitely large. The magnitude of the constant $\sigma_0^2$, however, cannot be greater than ${1\over 3}$ for positive values of matter density $\rho$. On the other hand, as $t\rightarrow \infty$, all the quantities $\rho,~\theta,~\sigma^2$ reduce to vanishingly small quantities. \\

It should be mentioned at this point that in view of all the models discussed so far, one can conclude that the Hawking-Penrose energy condition $(R_{ij} v^i v^j \le 0)$ is satisfied so long as $\rho \ge 0$. This is clear from the relation \eqref{2.23} and the fact that in the above models we have assumed $\zeta = 0$.\\

\noindent
\textbf{Case - 2:}\\
Let us now turn our attention to the nonvanishing values of bulk viscosity coefficients, that is, the situation where the bulk viscosity of the fluid cannot be completely ignored. We therefore assume
\be\label{3.20} \zeta = \zeta_0~~~~~\mathrm{and}~~~~~\eta = \eta_0 \rho^q,\ee
where, $\zeta_0$ and $\eta_0$ are constants. Now, integrating Eq. \eqref{3.2} one gets
\be\label{3.21} \theta = \left(\theta_0\over R^6\right) e^{{3\over 2}\zeta_0 t},\ee
$\theta_0$ being the constant of integration. Remembering that for the Bianchi I metric $\theta = 3\left(\dot R\over R\right)$ and so integrating further
Eq. \eqref{3.21}, we have the solution for $R$, which is given by
\be\label{3.22} R^3 = {\left(2\theta_0\over \zeta_0\right)}\left(e^{{3\over 2}\zeta_0 t} + D\right),\ee
where, $D$ is a constant of integration. Equation \eqref{3.20} therefore yields
\be \label{3.23} \theta = {3\over 2} \zeta_0\left[{e^{{3\over 2}\zeta_0 t}\over e^{{3\over 2}\zeta_0 t} + D}\right].\ee
Now from Eqs. \eqref{2.25} and \eqref{3.20} one gets
\be {\dot y\over y} = - \left(3\zeta_0 + 4\eta_0 \rho^q\right).\ee
which in turn can be written in view of Eq. \eqref{2.16} as
\be\label{3.24} {\dot y\over y} = -3\zeta_0 - 4\eta_0 \theta^{2q}\left({1\over 3} - y\right)^q.\ee
The special case for $D = 0$ is particularly simple and we discuss
only this case. We therefore have $\theta = {3\over 2}\zeta_0$ so that $\dot\theta = 0$. Here the expansion is steady and the rate is constant. One of
the relatively simple cases is $q = 1$. In this case $\eta = \eta_0\rho$ and
hence we obtain from Eq. \eqref{3.24}
\be\label{3.25} {\dot y\over y} = -3\zeta_0 - 4\eta_0 \theta^{2}\left({1\over 3} - y\right).\ee
Writing $3\zeta_0 = a_0$ and $9\eta_0^2 \zeta_0^2 = b_0$, the relation \eqref{3.25} may be written as
\be {\dot y\over y} = -a_0 -b_0\left({1\over 3} - y\right).\ee
where both $a_0$ and $b_0$ are greater than zero. Integrating Eq. \eqref{3.25} we further obtain
\be\label{3.26} {y\over (c_0 -y)} = e^{b_0c_0(t_0 - t)},\ee
So that one can write explicitly
\be\label{3.27} {\sigma^2\over \theta^2} = {c_0\over \left(1 + e^{b_0c_0(t - t_0)}\right)},\ee
where $c_0 = {a_0\over b_0} + {1\over 3}$, and $t_0$ is the constant of integration. Since here $\theta = {3\over 2} \zeta_0$ the expansion scalar $\theta$ is positive for the physical requirement $\zeta_0 > 0$. The above solution, therefore, represents an expanding model only. The density $\rho$ vanishes
at a finite time, when ${\sigma^2\over \theta^2} = {1\over 3}$, so that $\sigma^2$ remains finite. The proper volume represented by $R^3$ also has finite magnitude. But for $t \rightarrow \infty$, $R^3\rightarrow \infty,~\sigma^2 \rightarrow 0$, and $\rho\rightarrow {3\over 4}\zeta_0^2$. The singularity of vanishing volume $R^3 = 0$ exists at $t\rightarrow -\infty$, where the density is negative infinity. In fact prior to the instant corresponding to ${\sigma^2\over \theta^2} = {1\over 3}$, the density assumes only negative values. If one calculates $R_{ij}v^i v^j$ in this model, one finds it to be positive so that the energy condition is violated throughout.\\

Particularly, simple models can be constructed in this case, taking $q = 0$ in Eq. \eqref{3.20}, so that
\be\label{3.28} \zeta = \zeta_0,~~~~~\eta = \eta_0.\ee
The expressions for $R^3$ and $\theta$ remain unaltered from those given in Eqs. \eqref{3.22} and \eqref{3.23}, respectively. Equation \eqref{3.24}
then reduces to
\be\label{3.29} {\dot y\over y} = =(3\zeta_0 + 4\eta_0).\ee
This case is already mentioned in Eq. \eqref{2.34}. Integrating Eq. \eqref{3.29}, the solution can be obtained in the form
\be y = {\sigma^2\over \theta^2} = A^2 e^{-(3\zeta_0 + 4 \eta_0)t},\ee
so that
\be\label{3.30} \sigma^2 = {9\over 4} A^2 {\zeta_0}^2 e^{-4\eta_0 t}\left[e^{{3\over 2}\zeta_0 t} - D\right]^{-2},\ee
and the matter density $\rho$ is expressed as
\be \rho = \theta^2\left({1\over 3} - {\sigma^2\over \theta^2}\right) =
{9\over 4}\left[{{\zeta_0}^2 e^{3\zeta_0 t}\over \left(e^{{3\over 2}\zeta_0 t} + D\right)^2}\right]\left[{1\over 3} - A^2 e^{-(3\zeta_0 + 4\eta_0)t}\right].\ee
The maximum of ${\sigma^2\over \theta^2}$ is ${1\over 3}$ and this occurs at some finite
time $t$ given by Eq. \eqref{2.38}. In this limit $\rho = 0$ and $\sigma^2,~ \theta^2$ are both finite. For $D \ge 0$ the proper volume never reduces to zero magnitude as is evident from Eq. \eqref{3.22}. On the other hand, as $t\rightarrow \infty,~\theta \rightarrow {3\over 2}\zeta_0$, which is finite, $\sigma^2 \rightarrow 0$, and $\rho \rightarrow {3\over 2}{\zeta_0}^2$, but $R^3\rightarrow \infty$. We note that though the proper volume increases to large dimension and the anisotropy reduces to zero the fluid density $\rho$ does not vanish, unlike the Friedmann universe. For $D < 0$ we note that at a finite time $R^3 = 0$ and $\theta,~\sigma^2, ~\rho$ all become infinitely large. But, as $t\rightarrow \infty$, the proper volume also tends to infinity ($R^3 \rightarrow \infty$), while $\sigma^2 \rightarrow 0$ while, $\theta$ and $\rho$ both remain finite. In the limit $t\rightarrow \infty$ the behavior of the model is independent of the sign of the constant $D$.\\

For $D\ge 0$ it can be shown that the Hawking-Penrose energy conditions are violated throughout and for $D < 0$ this happens for large time $t$.

\section{Conclusion}

In summary, we have considered the Bianchi type I cosmological model with a viscous fluid, assuming the coefficients of viscosity as power functions of the matter density and considering an equation of state for a stiff fluid $(\rho = p)$. In most of the cases it has been observed that the matter density is zero at the initial singularity but then increases in the course of evolution, finally vanishing again in the asymptotic limit, which implies that the gravitational field creates matter. In the case of expanding models it is found that the dynamical importance of matter increases while that of shear decreases in the course of evolution. For a stiff fluid, in particular, this result is shown to hold also for contracting models. In addition to the role of shear viscosity in the dissipation of shear it is pointed out that the bulk viscosity can also augment the shear dissipation. For the stiff fluid with constant viscosity coefficients it is observed that the bulk viscosity can be an effective mechanism for large entropy in the asymptotic limit when the model approaches the isotropic Friedmann universe. Although the role of shear viscosity also is to increase the entropy in the course of the expansion, its effect becomes gradually less compared to that of the bulk viscosity in the asymptotic limit. The magnitude of the entropy is low at the highly anisotropic initial phase of evolution, as is observed from Eq. \eqref{2.37}, and then increases subsequently.\\

In the preceding section we have considered two different cases. The first case is $\eta = \eta_0\rho^n$ and $\zeta = 0$. Solutions for particular values of n such as $n = {1\over 2},~ 1, ~{3\over 2},~ 2$ are explicitly given. It is found that in all these cases except for $n = 1$, only expanding models are allowed. For $n = {1\over 2},~ 1, \mathrm{and} ~ 2$, the models have been found to approach the isotropic Friedmann universe asymptotically. The matter density vanishes at the initial phase of singularity, increases subsequently during the evolution, ultimately reduces to zero in the asymptotic limit. For $n = {3\over 2}$, the solution is not in the closed form and as such the behavior cannot be studied. For $n = 1$ as $t \rightarrow -\infty$, the proper volume $R^3$ tends to infinity whereas $\theta,~\sigma^2 ~\mathrm{and}~\rho $ vanish, representing a Friedmann model. In the course of time as $t \rightarrow 0$ the volume contracts and reduces to zero and the other physical scalars become infinitely large. In the case where the shear viscosity coefficient is assumed to be constant, i.e., $n = 0$, the behavior is different. Here at the initial phase of expansion $\theta,~\sigma^2 ~\mathrm{and}~\rho $ are infinitely large, but asymptotically, however, all of them reduce to vanishingly small quantities.\\

In the second case we have considered, $\eta = \eta_0\rho^q$, $\zeta = \zeta_0$· For $q = 0~ \mathrm{and}~ 1$, only expansion is found to be allowed. For $q = 1$, $\rho$ vanishes at a finite time, keeping the proper volume finite. In the limit $t\rightarrow \infty$, the model isotropizes with infinite proper volume but the matter density is non-vanishing, unlike the previous cases. For $q = 0$, either of the two cases is observed, depending on the sign of an integration constant $D$. For $D \ge 0$, the asymptotic behavior is the same as for $q = 1$. But for $D < 0$, the proper volume vanishes at a finite time while $\theta,~\sigma^2 ~\mathrm{and}~\rho $  take infinitely large magnitude at this instant. In the asymptotic limit, $t \rightarrow \infty$, the model is again identical to the case $q = 1$.

\end{document}